\documentclass[epj]{svjour}
\usepackage{amsmath}
\usepackage{siunitx}
\newcommand{\ww}[1]{\underline{\underline{{\bf #1}}}}
\newcommand{\tsigma}{\ww{\sigma}}

\newcommand{\fcap}{F^{\text{cap}}}

\newcommand{\be}{\begin{equation}}
\newcommand{\ee}{\end{equation}}
\newcommand{\bc}{\begin{center}}
\newcommand{\ec}{\end{center}}
\newcommand{\bea}{\begin{eqnarray}}
\newcommand{\eea}{\end{eqnarray}}
\newcommand{\ba}{\begin{aligned}}
\newcommand{\ea}{\end{aligned}}

\newcommand{\tr}{\text{tr}\hspace{.05em}}

\newcommand{\ave}[1]{\langle #1 \rangle}
\newcommand{\rij}{{\bf r}_{ij}}

\newcommand{\Fij}{{\bf F}_{ij}}

\newcommand{\scap}{\sigma^{\text{cap}}}
\newcommand{\scapc}{\sigma^{\text{cap,c}}}
\newcommand{\scapd}{\sigma^{\text{cap,d}}}
\newcommand{\fa}{\mathcal{F}}
\newcommand{\ld}{\mathcal{L}}
\newcommand{\fccis}{F^c_{12}}
\newcommand{\fdcis}{F^d_{12}}
\newcommand{\ffdcis}{\fa ^d_{12}}
\newcommand{\lldcis}{\ld ^d_{12}}
\newcommand{\scont}{\sigma^{\text{cont}}}

\newcommand{\tscap}{\tsigma^{\text{cap}}}
\newcommand{\tscont}{\tsigma^{\text{cont}}}
\newcommand{\tseff}{\tsigma^{\text{eff}}}

\newcommand{\Fn}{F^{\text{N}}}

\newcommand{\zc}{Z_{\text{C}}}
\newcommand{\zd}{Z_{\text{D}}}

\newcommand{\cP}{\mathcal{P}}
\newcommand{\cPcap}{\mathcal{P}^{\text{cap}}}

\newcommand{\mumc}{\mu^*_{\text{\tiny MC}}}
\usepackage{graphics}
\begin{document}
\title{Shear strength of wet granular materials: macroscopic cohesion and effective stress}
\subtitle{Discrete numerical simulations, confronted to experimental measurements }
\author{Michel Badetti\inst{1} \and Abdoulaye Fall\inst{1} \and Fran\c{c}ois Chevoir\inst{1} \and Jean-No{\"e}l Roux\inst{1}
}                     
%
%
\institute{Universit\'e Paris-Est, Laboratoire Navier, \\
IFSTTAR, ENPC, CNRS (UMR8205)\\
2 All\'ee Kepler, Cit\'e Descartes, F-77420 Champs-sur-Marne}
%
\date{Received: date / Revised version: date}
%
\abstract{Rheometric measurements on assemblies of wet  polystyrene bead assemblies, in steady uniform quasistatic shear flow, for varying liquid content within 
the small saturation (pendular) range of isolated liquid bridges, are supplemented with a systematic  study by discrete numerical simulations.
The numerical results agree quantitatively with the experimental ones provided that the intergranular friction coefficient is set to the value $\mu\simeq 0.09$, 
identified from the behaviour of the dry material. Shear resistance and solid fraction $\Phi_S$ are recorded as functions of the reduced pressure $P^*$, which, 
defined as $P^*=a^2\sigma_{22}/F_0$, compares stress $\sigma_{22}$, applied in the velocity gradient direction, to the tensile strength $F_0$ of the capillary bridges between grains of diameter $a$, and characterizes cohesion effects. 
The simplest Mohr-Coulomb relation with $P^*$-independent cohesion $c$ applies in good approximation for large enough $P^*$ (typically $P^*\ge 2$). 
Numerical simulations extend to different values of $\mu$ and, compared to experiments, to a wider range of $P^*$.
The assumption that capillary stresses  act similarly to externally applied ones onto the dry granular contact network  (effective stresses) leads to very good 
(although not exact) predictions of the shear strength, throughout the numerically investigated range $P^*\ge 0.5$ and $0.05\le\mu\le 0.25$. 
Thus, the internal friction coefficient $\mu^*_0$ of the dry material still relates the contact force contribution to stresses,
$\scont _{12}=\mu^*_0 \scont_{22}$, while the capillary force contribution to stresses, $\tscap$, defines a generalized Mohr-Coulomb cohesion $c$, 
depending on $P^*$ in general. $c$ relates to $\mu^*_0$, coordination numbers and  capillary force network anisotropy. 
$c$ increases with liquid content through the pendular regime interval, to a larger extent the smaller the friction coefficient. The simple approximation ignoring
capillary shear stress $\scap_{12}$ (referred to as the Rumpf formula) leads to correct approximations for the larger saturation range within the pendular regime, 
but fails to capture the decrease of cohesion for smaller liquid contents.}
%
\authorrunning{Badetti \emph{et al.}}
\titlerunning{Shear strength of wet granular materials}
\maketitle
\section{Introduction}
\label{intro}
\begin{figure}[h]
  \centering
  \resizebox{0.75\columnwidth}{!}{\includegraphics{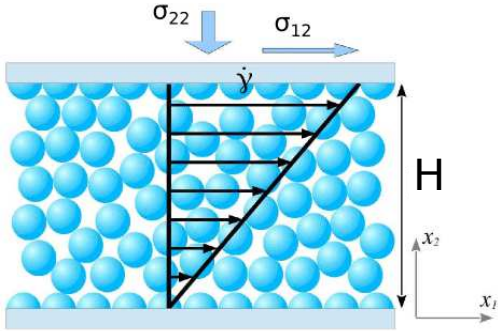}}
  \caption{Sketch of a shear test, in which the material flows along coordinate axis $x_1$, with controlled 
  normal stress $\sigma_{22}$ along the velocity gradient direction, coordinate axis $x_ 2$.  }
\label{fig:sheartest}
\end{figure}

Wet granular materials~\cite{MiNo06}, in which grains  are large enough for colloidal forces to be negligible, differ from dry ones by their \emph{cohesion}. On the grain scale, this cohesion is due to the pressure in the wetting liquid being lower than in the surrounding atmosphere, and
thereby effectively attracting the wet grains to one another. For low liquid contents, the wetting liquid, in the so-called pendular regime, forms isolated
bridges joining pairs of grains in contact or separated by a short distance. Those liquid bridges transmit an attractive capillary force, depending on
grain geometry, bridge volume and interfacial tension. The value $F_0$ of this attractive force for contacting grains sets a  force scale, which 
micromechanical models of dry granular materials are usually devoid of. Similarly, at the macroscopic scale, a characteristic
stress value $c$ is introduced into the plastic  flow criterion, often assumed, for shear flow in direction 1, 
with velocity gradient in direction 2  (see Fig.~\ref{fig:sheartest}), to
take the celebrated Mohr-Coulomb form relating shear stress $\sigma_{12}$ to normal stress $\sigma_{22}$ in the gradient direction as
\be
\vert\sigma_{12}\vert = \mumc \sigma_{22} + c,
\label{eq:mohrcoulomb}
\ee
in which $\mumc$ is the internal friction coefficient and $c$, which vanishes for cohesionless materials such as dry sands,  is referred to the macroscopic material cohesion. [The absolute value written in \eqref{eq:mohrcoulomb} is due to our sign convention for the stress tensor: 
compressive stresses are positive, and 
one has $\sigma_{12}<0$ in a simple shear flow with velocity field $v_1$ in direction 1, with gradient $\partial v_1/\partial x_2 >0$ (Fig.~\ref{fig:sheartest})]. 
Over the past decades, some experimental studies~\cite{PiCa97,GTH03,PAC98,RYR06} and a larger number of numerical ones~\cite{GTH03,RYR06,Sou06b,SCND09,SHNCD09,KRC15,SJT16}  have been carried out on wet spherical beads, 
as a model material, to investigate  macroscopic cohesion values and explore, in particular, the relation between $F_0$ and $c$. 
Visualizations of liquid bridge networks in such materials have been rendered possible in the lab by cleverly designed techniques~\cite{KOH04,HER05}. 
Numerical simulations investigate grain-level origins of macroscopic mechanics through the ``discrete-element method" (DEM), the granular material analog of molecular dynamics~\cite{RD11}. DEM is directly applicable to wet grains in the pendular regime,
for which pairwise additive capillary forces have been suitably modeled~\cite{LiThAd93,WAJS00,PMC00,HER05}.

 The idea that capillary forces, by pressing neighbouring grains onto one another, act similarly to an
external isotropic pressure applied to the pack of grains leads to the following simple relation, 
between $F_0$ and $c$, 
involving solid fraction $\Phi_S$, grain diameter $a$ and coordination number $Z$ of liquid bridges, and the Mohr-Coulomb internal friction coefficient:
\be
c = \mumc\frac{Z\Phi_SF_0}{\pi a^2}.
\label{eq:Rumpf}
\ee
This relation, to which we refer as the Rumpf formula (as it is often attributed to Ref.~\cite{Rumpf70}),   was discussed, reestablished or reformulated in many publications~\cite{PAC98,GTH03,RYR06,KRC15}, which often found it to
provide rather good estimates of the macroscopic cohesion~\cite{KRC15}.

The assumption that capillary forces have an effect equivalent to an additional isotropic pressure on the cohesionless grains has nevertheless been criticized, 
on the ground that the capillary stress on the granular network is not isotropic in conditions of
macroscopic yield under shear~\cite{SCND09,SHNCD09,RR09,SJT16,Chareyre}.  
Those references pointed out that, if the yield condition of the wet material is to be likened to the 
yield condition of the dry one under some modified, \emph{effective} stress tensor $\tseff$, then the capillary tensor, $\tscap$ which defines it at 
$\tseff=\ww{\sigma}-\tscap$, is not isotropic and has a non-negligible shear component $\scap_{12}$. 

These issues are revisited in the present paper, which deals, experimentally and numerically,  with  wet spherical bead assemblies. 
We investigate whether  wet bead assemblies satisfy the Mohr-Coulomb relation~\eqref{eq:mohrcoulomb}.
We use both experiments and DEM simulations, for which the system choice and experimental or numerical setups are described in Sec.~\ref{sec:meth}. 
Laboratory experiments are used  to quantitatively validate the numerical simulations, provided the basic material characteristics are correctly identified. Such identification is carried out in Sec.~\ref{sec:calibsec}, based on
the properties of dry grains in shear flow. In Sec.~\ref{sec:mohrcoulomb}, the applicability of the Mohr-Coulomb criterion is discussed and macroscopic 
cohesion $c$ is measured for the laboratory system, as well as its numerical counterpart. 
DEM computations are then used, in Sec.~\ref{sec:effective}, to  study  the influence of material parameters (intergranular friction coefficient, interfacial tension of the wetting liquid, liquid content) onto macroscopic properties $\mumc$ and $c$. 
Over the explored range of state parameters, we investigate the possible relation, through the definition of some effective stress, of the yield condition of the wet material in quasistatic shear  to the one of the dry grains, and 
relate the macroscopic shear resistance to micromechanical and microstructural variables.  Sec.~\ref{sec:s12cap} is devoted to the micromechanical origins of the contribution of capillary forces to shear stress, $\scap_{12}$. 
In the final part, Sec.~\ref{sec:conc}, the results are summarized and put in perspective.

 \section{Methods and parameters\label{sec:meth}}
 \subsection{Experimental}
\begin{figure}[h]
  \centering
   \resizebox{0.75\columnwidth}{!}{\includegraphics{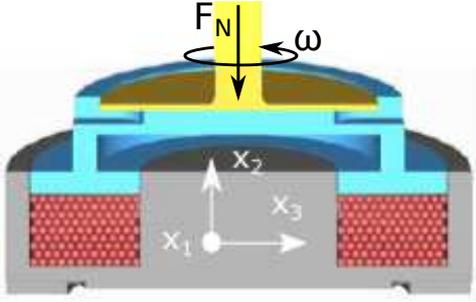}}
  \caption{Rheometric cell used in experiments, seen as cut by azimuthal plane containing axis of rotation.
\label{fig:rheometer}}
\end{figure}

 We consider assemblies of spherical polystyrene beads, with some narrow diameter distribution about the mean value $a=\SI{500}{\micro\m}$, wet by a non-volatile liquid, a silicone oil (47V50 provided by Chem+), with interfacial tension $\Gamma =\SI{20.6}{\milli\N\per\m}$, and small wetting angle $\SI{2}{\degree}\le\theta\le\SI{5}{\degree} $.
 A fixed (small) amount of liquid is first mixed with the beads. We denote as $\Phi_L$ and $\Phi_S$ the volume fractions of the liquid and of the solid beads in the system. Since the sample is allowed to dilate, both volume fractions vary. 
 We refer to their fixed ratio, $\Phi_L/\Phi_S$ as the  liquid content. The saturation, $S$, is given by
 $$
 S = \frac{\Phi_L}{1-\Phi_S}.
 $$
 The wet system is then placed inside a rotative annular rheometer as represented in Fig.~\ref{fig:rheometer}. This apparatus is the same as the one used in Ref.~\cite{FOHMRC15}, which can be consulted for more details.
The annulus-shaped cell containing the grains is limited by smooth cylindrical walls in the radial direction, and by a rough bottom surface underneath.  
It is closed on top by the inferior rough surface of a mobile lid, which moves in rotation about the axis at a prescribed angular velocity $\omega$, while its vertical position (the coordinate $x_2$ of the cell
 ``ceiling'' in Fig.~\ref{fig:rheometer}, or the cell height $H$) is free to adjust in order to exert a controlled vertical force $F_N$ onto the granular sample. We measure the torque, $T$, that is necessary to maintain a fixed, prescribed value of $\omega$, in the steady state, as well as
 the sample height $H$. Denoting as $R_i$ and $R_e$ the inner and the outer radius of the annular cell, and as $V_S$ the total volume of the beads, control parameters 
 $\omega$ and $F_N$, and measured quantities $T$ and $H$ are translated into more intrinsic rheological terms using the following relations:
 \be
 \ba
 \sigma_{22} & \simeq \frac{F_N}{\pi R_e^2\left[ 1-\frac{R_i^2}{R_e^2}\right]}\\
  \sigma_{12} & \simeq \frac{3T}{\pi R_e^3\left[ 1-\frac{R_i^3}{R_e^3}\right]}\\
\Phi_S&=\frac{V_S}{\pi H R_e^2\left[ 1-\frac{R_i^2}{R_e^2}\right]}\\
\dot\gamma &\simeq \omega\frac{R_i+R_e}{2H}. 
 \ea
 \label{eq:exprheol}
 \ee
 Measurements are carried out in the steady state, as torque $T$ and height $H$ remain constant. 
 The material being continuously sheared has then reached constant solid fraction $\Phi_S$ and shear stress $\sigma_{12}$. 
 Provided the material state is uniform (which may be checked by comparison with numerical simulations, or on investigating possible size effects on the measured rheology), such steady states are independent of initial conditions. In the quasistatic limit of low shear rate $\dot\gamma$, the experiment probes the classical ``critical state" 
of monotonic simple shear,  as defined in soil mechanics~\cite{DMWood,AFP13}.

 \subsection{Numerical\label{sec:num}}
We carry out DEM simulations with the same model as in Refs.~\cite{KRC15,TKTPCR17}. Monosized beads of diameter $a$ interact at their contacts by elasticity and friction, which are modeled through a standard simplified form of
Hertz-Mindlin laws~\cite{iviso1}, involving the elastic properties of the  material which the beads are made of, and an intergranular friction coefficient we simply denote as $\mu$. 
While contact elasticity is irrelevant to the granular material rheology in the investigated range~\cite{RoCh11} (it would, of course, matter, should we become interested in the elastic properties of the granular assembly), 
$\mu$ is an important parameter, the appropriate value of which for the experiments  is identified below in Sec.~\ref{sec:calibsec}. 

\begin{figure}[h]
  \centering
     \resizebox{0.75\columnwidth}{!}{\includegraphics{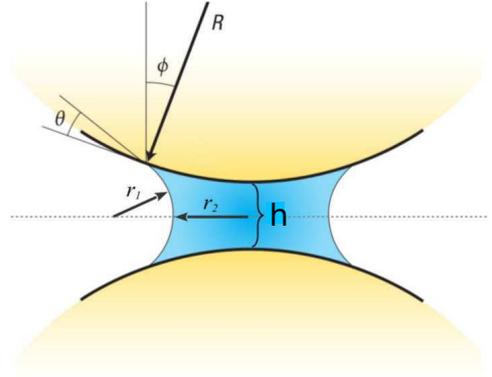}}
  \caption{Liquid bridge joining two spherical beads.
\label{fig:bridge}}
\end{figure}

We consider small liquid contents, and restrict the model to the pendular regime in which the liquid forms disjoint menisci bridging
 pairs of grains in contact or close to one another (Fig.~\ref{fig:bridge}).
As a parameter of the numerical model, we fix the meniscus volume $V_0$. 
The meniscus is supposed to form, out of the small quantity of liquid coating the small asperities on the grain surfaces, 
as soon as two grains come into contact. If the grains subsequently move apart from each other,  
the meniscus deforms and breaks when the  distance between their surfaces, $h$, reaches the rupture threshold~\cite{LTA93}, 
\be
D_0 = V_0^{1/3}.
\label{eq:drupt}
\ee
The liquid bridge, when present, introduces an attractive capillary force for which we adopt the simple Maugis model~\cite{Maugis87}, 
as in  Refs.~\cite{KRC15,TKTPCR17}. This model is suitable for small enough $V_0$, as tested in Ref.~\cite{PMC00}\footnote{With the notations of Fig.~\ref{fig:bridge}, the Maugis approximation assumes $r_1\ll r_2$ and very small filling angle $\phi$.}.
The maximum attractive force (tensile strength), reached for contacting particles, 
is independent of $V_0$ and equal to  $F_0 = \pi a \Gamma$, $\Gamma$ denoting the liquid interfacial tension, if perfect wetting is assumed (otherwise 
one should multiply $\Gamma$ by  the cosine of the wetting angle). 
 The  capillary force varies with distance $h$ between particle surfaces as
\begin{equation}
   \fcap =
   \begin{cases}
      -F_0 & \mbox{if}\ h \leq 0\ \ \mbox{(contact)}\\
      -F_0\left[1-\dfrac{1}{\sqrt{1+\dfrac{2V_0}{\pi a h^2}}}\right] & \mbox{if } 0 < h \leq D_0,
         \end{cases}
\label{eq:Maugis}
\end{equation}
and is only present, for noncontacting grains ($h>0$), if they have  been in contact in the past and have never been separated by a distance exceeding $D_0$  since. 

The simplifying assumptions adopted for the distribution of liquid in the mixture entail that the conditions of conservation of liquid volume (for non-volatile liquids) 
or the capillary pressure uniformity (for liquids in equilibrium with their vapour) are 
not satisfied. The  consequences of this minor drawback of the model, as assessed in~\cite{KRC15} are however quite innocuous, because moderate changes in liquid contents, associated with the variations of the number of liquid bonds throughout the
investigated states, have negligible effects on rheological properties. Liquid contents as expressed by ratio $\Phi_L/\Phi_S$ have to be measured in simulations, depending on $V_0$ and on the coordination number $Z$  of the liquid bridge network (average
number of bridges connecting  one grain to its neighbours):
\be
\frac{\Phi_L}{\Phi_S}= \frac{3Z V_0}{\pi a^3}
\label{eq:lcont}
\ee

Assemblies of 4000 spherical beads are placed in a cuboidal cell, periodic in all three directions, with an adjustable height $H$, so that a constant stress $\sigma_{22}$ is maintained, 
while the Lees-Edwards method is implemented~\cite{KRC15} to impose
velocity gradient (shear rate) $\dot\gamma = \dfrac{\partial v_1}{\partial x_2}$ at the macroscopic scale (see, e.g., \cite{PR08a} for more details on these manipulations  of  boundary conditions, enforcing normal stress-controlled shear flows within periodic cells). 

Shear stress $\sigma_{12}$ and solid fraction $\Phi_S$ are measured in steady state flow. 
Stress components $\sigma_{\alpha\beta}$ 
are evaluated with the usual formula (see e.g. ~\cite{CMNN81}),
as a  sum, divided by sample volume $\Omega$, 
over interacting grain pairs $i$--$j$, involving the force, $\Fij$, 
transmitted from grain $i$ to grain $j$ in their contact or through a small liquid bridge, and 
the center-to-center vector, $\rij$, pointing from $i$ to $j$:
\be
\sigma_{\alpha\beta}= \frac{1}{\Omega}
\sum_{i<j} F_{ij}^{\alpha} r_{ij}^{\beta}
\label{eq:stress}
\ee
[Away from the quasistatic limit a kinetic term  should be added in \eqref{eq:stress}].
One may separate the capillary forces from the contact ones in the sum of Eq.~\ref{eq:stress}, and accordingly define capillary and contact contributions to stresses:
 \be
\sigma_{\alpha\beta}= \scont_{\alpha\beta}+\scap_{\alpha\beta}.
\label{eq:stressdec}
\ee
As noted in previous publications~\cite{KRC15}, the average pressure, $\cP=\tr \ww{\sigma}/3$ is related to the average normal force 
in all interacting pairs, $\langle \Fn\rangle$, and to the average, $\langle \Fn h\rangle _d $, over capillary forces attracting non-contacting pairs through a liquid bridge, 
of the product of force by distance $h\le D_0$:
\be
  \cP = {\Phi Z \over \pi a^2} \langle \Fn\rangle + {\Phi \zd \over \pi a^3} \langle \Fn h \rangle _d,
\label{eq:PvsFN}
\ee
where $\zd$ denotes the   \emph{coordination number of distant interactions}.
In formula~\eqref{eq:PvsFN}  the second term contributes at most 2\% of the pressure. 
Such a relation holds separately for contact and capillary forces, and yields in the latter case 
(keeping notation $\langle...\rangle_d$ for averages over pairs interacting through a meniscus without contact)
\be
  \cPcap =- {\Phi \zc \over \pi a^2} F_0 + {\Phi \zd \over \pi a^2} \langle \fcap(h)  \rangle _d+{\Phi \zd \over \pi a^3} \langle \fcap(h) h \rangle _d
\label{eq:PcapvsFN}
\ee
In \eqref{eq:PcapvsFN} we have introduced notations $\zc$ for the \emph{contact coordination  number} (the total coordination number is $Z=\zd+\zc$). 
The capillary forces depending on gap $h$ appearing in the second and third terms, from \eqref{eq:Maugis}, 
are negative with intensity lower than $F_0$. We define $F_d$ ($0\le F_d\le F_0$) by: 
\be
-F_d= \langle \fcap(h)  \rangle _d
\label{eq:deffd}
\ee
The last term of \eqref{eq:PcapvsFN} is negligible. 

\subsection{Dimensionless control parameters}
In addition to the material parameter $\mu$ and to liquid content $\Phi_L/\Phi_S$, the state of the  material, in simple shear flow with strain rate $\dot\gamma$,  depends on two important dimensionless parameters. The first one is the
\emph{inertial number}, as used in many rheological studies of dry and wet granular materials~\cite{Gdr04,Dacruz05,JFP06,Hatano07,RoRoWoNaCh06,PR08a,Boyer2011,AFP13,AzRa2014,KRC15,FOHMRC15,BeAzDoRa16} 
($m$ denotes the mass of one grain):
\be
I = a\dot\gamma\sqrt{\frac{m}{\sigma_{22}}}.
\label{eq:defI}
\ee
We are mostly interested in the quasistatic limit of $I\to 0$, which is approached with good accuracy, with frictional grains, for $I\sim 10^{-3}$. 

In the presence of adhesive forces, the second, important dimensionless control parameter is the reduced pressure, $P^*$~\cite{KBBW03,GiRoCa08,RoCh11,KRC15} comparing the characteristic force $F_0$ (adhesive strength) 
to the controlled normal stress $\sigma_{22}$: (some authors~\cite{RoRoWoNaCh06,BeAzDoRa16} use a ``cohesion number'' defined as $\eta = 1/P^*$)
\be
P^* = \frac{a^2\sigma_{22}}{F_0}.
\label{eq:defPstar}
\ee
Adhesive forces dominate for small $P^*$ and tend to stabilize loose structures, either in static packs~\cite{KBBW03,GiRoCa08,TKTPCR17}, or in shear flow~\cite{RoRoWoNaCh06,KRC15,BeAzDoRa16}. 
For large $P^*$, adhesive forces become negligible and the properties of dry, cohesionless grains are retrieved. In the experiments reported here, $P^*=1$ corresponds to $\sigma_{22}=\SI{0.129}{\kilo\Pa}$.

As in previous experimental~\cite{FOHMRC15} and numerical~\cite{Dacruz05,KRC15} studies, 
we shall simply denote as $\mu^*$ the apparent (secant) internal friction coefficient, defined
as the ratio of the shear stress to the normal stress:
\be
\mu^* = \frac{\vert\sigma_{12}\vert}{\sigma_{22}}.
\label{eq:defmustar}
\ee
$\mu^*$ depends on inertial number $I$ and on $P^*$ in general~\cite{AFP13,FOHMRC15,KRC15}. In the present study, we only consider the quasistatic limit of
$I\to 0$. We denote as $\mu^*_0$ the quasistatic  value of $\mu^*$ in the dry material ($\Phi_L=0$). $\mu^*$ takes value 
$\mu^*_0$ in the double limit of  $I\to 0$  and large $P^*$.
\section{Calibration of DEM simulations with the experimental data: dry grains\label{sec:calibsec}}
In this section we confront experimental measurements and numerical simulations for the internal friction coefficient of the material in shear flow, in the case of dry grains, which are known to exhibit a well-defined internal friction coefficient, but no 
macroscopic cohesion: relation~\eqref{eq:mohrcoulomb} is well satisfied with a finite 
$\mumc=\mu^*_0$ and $c=0$ [$\mumc$ is the $\sigma_{22}$-independent value of 
$\mu^*$ in \eqref{eq:defmustar}].
Experiments, carried out for $\SI{0.3}{\kilo\Pa}\le\sigma_{22}\le \SI{3}{\kilo\Pa}$, yield $\mu^*_0 = 0.25\pm 0.02$ in slow, steady shear flow, while 
the system solid fraction stabilizes at $\Phi_S = 0.615\pm 0.002$. 

\begin{figure}[h]
  \centering
    \resizebox{0.9\columnwidth}{!}{\includegraphics{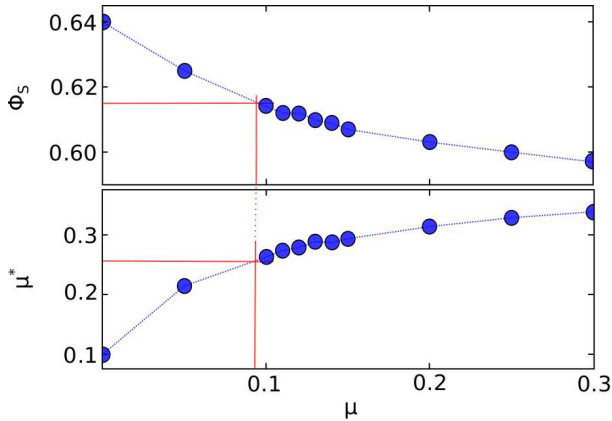}}
  \caption{Internal friction coefficient $\mu^*$ and solid fraction $\Phi_S$ measured in numerically simulated steady quasistatic shear flow of dry grains, 
  versus intergranular friction coefficient $\mu$. Values obtained for $\mu=0.09$ are shown with red lines.
\label{fig:mustarphimu}}
\end{figure}

In the numerical simulations, both  $\mu^*_0 = \vert\sigma_{12}\vert/\sigma_{22}$ and  $\Phi_S$, as observed in quasistatic flow,  depend on intergranular friction coefficient $\mu$, as shown in Fig.~\ref{fig:mustarphimu}.
As previously reported~\cite{LRC09}, $\Phi_S$ is a decreasing function of $\mu$, while $\mu^*$ increases, both starting at the well-defined values for $\mu=0$~\cite{PR08a}, i.e., the ``random close packing''  solid fraction near $0.64$,  
and the  internal friction coefficient of assemblies of frictionless beads, $\mu^*\simeq 0.1$. 

Remarkably, both quantities nearly simultaneously reproduce the experimental results for $\mu=0.09$: numerical values are then $\mu^*_0= 0.257\pm 0.002$ and $\Phi_S= 0.6150\pm \num{3e-4}$.
This value $\mu=0.09$ of the intergranular friction coefficient is thus adopted in the following in order to simulate the material tested in the laboratory.

\begin{figure}[h]
  \centering
(a)\resizebox{0.9\columnwidth}{!}{\includegraphics{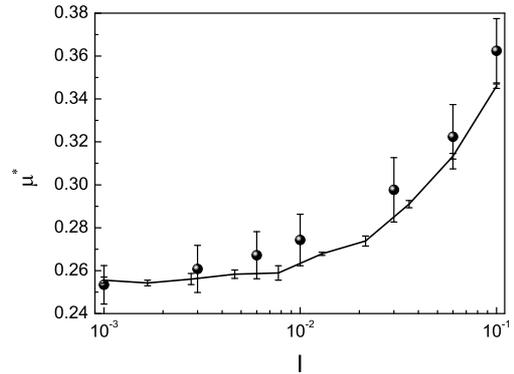}}
(b)\resizebox{0.9\columnwidth}{!}{\includegraphics{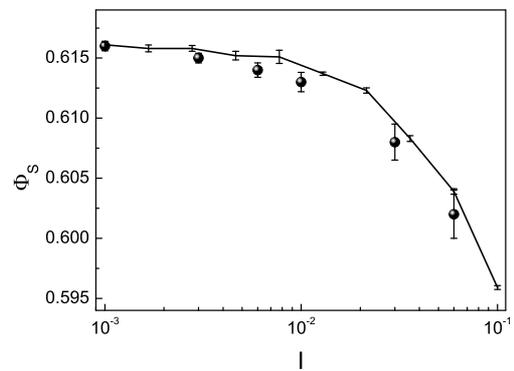}}
  \caption{$I$-dependent internal friction coefficient $\mu^*$ (a) and solid fraction $\Phi_S$ (b), for both laboratory (round dots) and  numerical results (data points joined by continuous line), computed with $\mu=0.09$. 
\label{fig:muphinertial}}
\end{figure}

Furthermore, the quantitative agreement between numerical and experimental results is observed to extend to flows with inertial effets: 
the $I$ dependence of both quantities $\mu^*$ and  $\Phi_S$, as observed in the laboratory, is well reproduced by the numerical simulations, as apparent in Fig.~\ref{fig:muphinertial}.
We are thus in a good position to confront numerical results to experimental ones in the case of wet grains, and to discuss the definition of cohesion $c$ and the influence of $P^*$ and $\Phi_L/\Phi_S$ on the material behaviour.
\section{Mohr-Coulomb cohesion: experimental and numerical results\label{sec:mohrcoulomb}}
Once the intergranular friction coefficient is identified from the rheology of dry bead assemblies, assuming the value of $\mu$ does not change in the presence of the wetting liquid, 
we turn to the identification of $\mu^*$ and $c$, the parameters of a Mohr-Coulomb relation applying to the chosen material.

Classically, these parameters are measured on fitting a straight line through the values of $\sigma_{12}$ plotted versus $\sigma_{22}$. 
Alternatively, it might prove convenient to search for a normalized, dimensionless cohesion, defined as
\be
c^* = \frac{a^2c}{F_0},
\label{eq:defcstar}
\ee
and to study the variations, at small enough $I$,  
of stress ratio $\mu^*$ (see Eq.~\ref{eq:defmustar}) versus $P^*$, thereby reformulating the Mohr-Coulomb relation as
\be
\frac{\vert\sigma_{12}\vert}{\sigma_{22}} = \mumc + \frac{c^*}{P^*}.
\label{eq:mohrcoulomb2}
\ee
One may thus seek a linear variation of  stress ratio $\mu^*$ with $1/P^*$. 

For large $P^*$, adhesive capillary forces become negligible, and the material behaviour should be the same as in the absence of the wetting fluid. If a Mohr-Coulomb criterion applies, then parameter $\mumc$ necessarily coincides with the
internal friction coefficient of the dry material, $\mu^*_0$ (equal to $0.25\pm 0.02$ in the present case).

\subsection{Parameter range\label{sec:range}}
In experiments, practical limitations apply to the values of $P^*$. First, stresses within the sample should remain reasonably homogeneous. Vertical stress
$\sigma_{22}$ varies through the thickness of the sample, due to the weight of the grains, and the results are given as functions of the average value, at mid-height. 
In practice we do not often record rheological characteristics for $P^*$ below 2.3 ($\sigma_{22} = \SI{0.3}{\kilo\Pa}$), which correspond to relative stress variations $(\Delta\sigma_{22})/\sigma_{22}$ above $\num{5}$\% within the sample.
Furthermore, it proves difficult in practice to control low levels of stress. Thus states with $P^*=1.55$ and $P^*=0.773$ are only satisfactorily obtained for the largest liquid content, $\Phi_L/\Phi_S=0.075$ (with, however, $(\Delta\sigma_{22})/\sigma_{22}$ reaching $0.2$). 
Simulation results extend down to $P^* = 0.5$, but smaller confining stresses make it difficult to observe  uniform flows,  due to strong strain localization tendencies, 
as reported in Ref.~\cite{KRC15}. This earlier work introduced localization
indicator  $\Delta$, computed from velocity profiles $v_1(x_2)$ across the cell in the gradient direction (for $-H/2\le x_2\le H/2$) as 
\be
\Delta = \frac{12}{H^3\dot\gamma^2}\int_{-H/2}^{H/2}\left[ v_1(x_2)-\dot\gamma x_2\right]^2dx_2.
\label{eq:defdelta}
\ee
Shear flow and boundary conditions are enforced such that $\Delta$ vanishes if the shear rate is uniform, and
 $\Delta$ is equal to 1 if the case of an infinitely thin shear band between two solid blocks sliding on each other. We checked that its time average did not exceed 0.05 for all $P^*\ge 0.5$, in agreement with~\cite{KRC15} (with fluctuations of order 0.1 for $P^*=0.5$, the case with the larger departures from the average linear velocity profile).
 
\begin{figure}[h]
  \centering
     \resizebox{0.9\columnwidth}{!}{\includegraphics{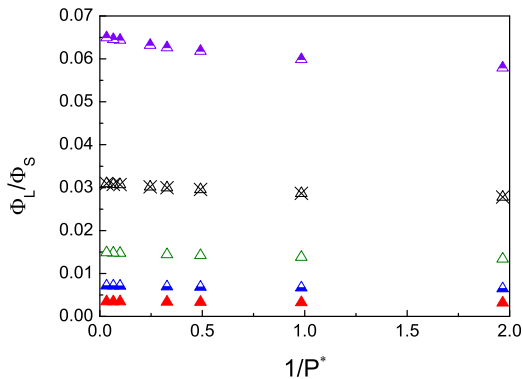}}
  \caption{Liquid content $\Phi_L/\Phi_S$ versus $P^*$ for the 5 different values of $V_0/a^3$ used in the simulations (see Table~\ref{tab:philphis}).
\label{fig:philphis}}
\end{figure}

Liquid content $\Phi_L/\Phi_S$, as noted in Sec.~\ref{sec:num}, is not rigourously constant if $V_0$ is kept fixed while normal stress varies, but, as shown in Fig.~\ref{fig:philphis}, its variations are not really significant.
The correspondence between liquid content and meniscus volume is given by Table~\ref{tab:philphis}.
\begin{table}[h!]
\centering
\caption{Approximate correspondence (see Fig.~\ref{fig:philphis} -- we use the value of the liquid content corresponding to $P^*=3$) between prescribed meniscus volume and liquid content.}
\label{tab:philphis}       
\begin{tabular}{|c||c|c|c|c|c|}
\hline\noalign{\smallskip}
$V_0/a^3$ & $5.10^{-4}$ & $10^{-3}$ & $2.10^{-3}$& $4.10^{-3}$& $8.10^{-3}$\\
\noalign{\smallskip}\hline\noalign{\smallskip}
$\Phi_L/\Phi_S$ & $0.003$ & $0.007$ & $0.014$& $0.030$& $0.063$\\
\noalign{\smallskip}\hline
\end{tabular}
\end{table}
The meniscus volume associated with the upper limit of the pendular range, reached as the menisci joining three beads, 
each one in contact with the other two, start to merge, is about $8.10^{-3}a^3$~\cite{KRC15}, from which we set the maximum value in Table~\ref{tab:philphis}. 
Laboratory measurements cover roughly the same range of liquid contents.

\subsection{Measurement of a macroscopic cohesion\label{sec:cexpnum}}
\begin{figure}[htb!]
\centering
\resizebox{1.15\columnwidth}{!}{\includegraphics{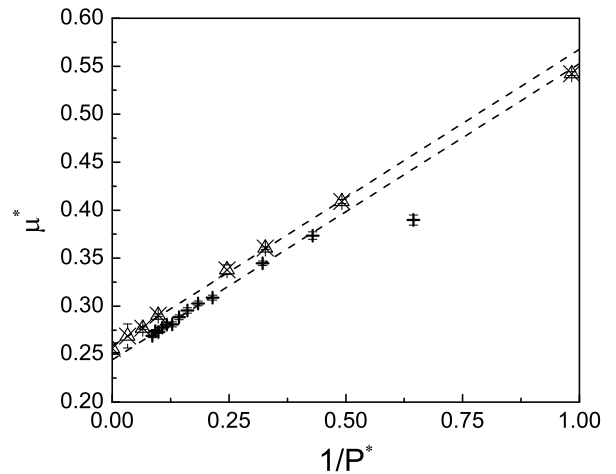}}\\
\caption{ Experimental (+ symbols) and numerical (crossed hollow triangles) results for stress ratio $\vert\sigma_{12}\vert/\sigma_{22}$ versus $1/P^*$ 
for $\Phi_L/\Phi_S=0.03$. 
Data fitted with Eq.~\ref{eq:mohrcoulomb2} (dotted lines) for $P^*\ge 2.3$. 
\label{fig:cexpnum}}
\end{figure}

\begin{figure}[htb!]
\centering
\resizebox{1.05\columnwidth}{!}{\includegraphics{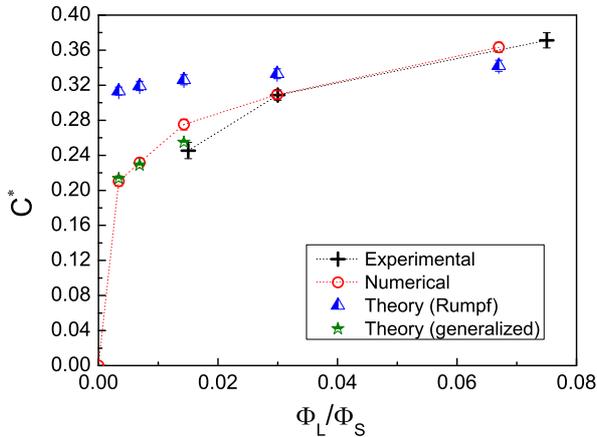}}\\
\caption{Macroscopic (reduced, dimensionless) cohesion $c^*$ (in units of $F_0/a^2$) versus liquid content $\Phi_L/\Phi_S$, as measured 
in experiments and in numerical simulations, and as predicted by the Rumpf expression~\eqref{eq:Rumpf} or by a more elaborate approach (presented in 
 Sec.~\ref{sec:effective}).
\label{fig:cphilphis}}
\end{figure}

Fig.~\ref{fig:cexpnum}  shows how  linear relation \eqref{eq:mohrcoulomb2} may be used to identify values of cohesion $c$ 
(or $c^*$, see Eq.~\ref{eq:defcstar}), for both experimental and numerical results. 
Eq.~\ref{eq:mohrcoulomb2} is a successful fit to the data for $P^*\ge 2.3$. The straight lines intercept the vertical axis at nearly the same values for experimental and numerical data, 
both confirming $\mumc=\mu^*_0$ (the data appear to indicate a very slightly smaller friction coefficient in the experiment). 
The figure also illustrates the good agreement between simulations and experiments for reduced cohesions $c^*$: both straight lines are parallel.
For smaller $P^*$, the shear resistance is overestimated by the Mohr-Coulomb criterion fitted to larger $P^*$ values. For $P^*\ge 1$ 
the discrepancy is apparently larger with experimental results, but
we already pointed out (Sec.~\ref{sec:range}) that laboratory measurements are more problematic in that range. 
On identifying (dimensionless) cohesion $c^*$ for varying liquid contents by the same fitting procedure, 
one obtains the cohesion values shown in Fig.~\ref{fig:cphilphis}.  
This figure  shows that the macroscopic cohesion is a growing function of the liquid content.
The agreement between numerical and experimental results is quite satisfactory. The ``Rumpf formula" (Eq.~\ref{eq:Rumpf})  correctly predicts  macroscopic cohesion $c$ for the largest values of liquid content in the investigated range, but overestimates it at lower liquid contents, for which it fails 
to capture the decreasing trend. Note that some constant 
value of the coordination number has to be chosen in~\eqref{eq:Rumpf} -- the choice made for the data points shown in Fig.~\ref{fig:cphilphis} is explained and
discussed in Sec.~\ref{sec:effective}. In Sec.~\ref{sec:effective}, we also obtain the more sophisticated estimate of $c^*$, shown in Fig.~\ref{fig:cphilphis} as star-shaped data points,
which comes closer to the experimental and numerical values for small $\Phi_L$.

More complete experimental results are presented in another paper~\cite{Abdou2}, in which inertial flows, departing from the quasistatic limit,  
are also considered. 
\subsection{Solid fraction}
We report here on the variations of  solid fraction $\Phi_S$, from both experiments and simulations, 
with reduced pressure $P^*$ and liquid content $\Phi_L/\Phi_S$. 
\begin{figure}[htb!]
\centering
\resizebox{0.92\columnwidth}{!}{\includegraphics{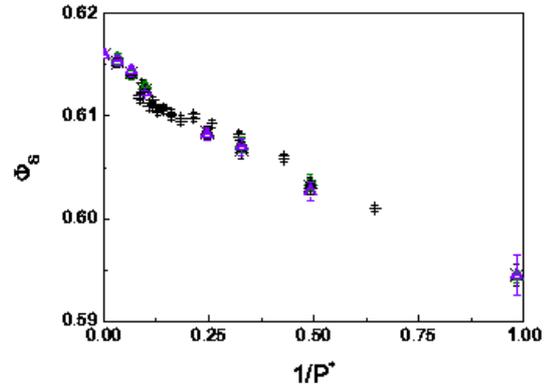}}\\
\caption{Solid fraction $\Phi_S$ versus $1/P^*$. Experimental (+ symbols) and 
numerical results (triangles) are shown for different liquid contents $\Phi_L/\Phi_S$.
\label{fig:phiexpnum}}
\end{figure}
Fig.~\ref{fig:phiexpnum} shows that the system density in steady quasistatic  shear flow tends to decrease for growing $1/P^*$, although it is  much less sensitive
to $P^*$  than the shear resistance,  displayed in Fig.~\ref{fig:cexpnum}.
Capillary forces entail a decrease, from the dry value $\simeq 0.615$ in the limit of large $P^*$ 
down to  $\Phi_S\simeq 0.594$ at $P^*=1$. Numerical results quantitatively agree with experimental ones when both are available 
and the solid fraction is not influenced by the liquid content.
\section{Cohesion and effective stress: a numerical study\label{sec:effective}}
Having shown that numerical simulations nicely agree with laboratory results in the range of parameters accessible to the experiments, 
we now use the numerical tool to further investigate the shear resistance-enhancing effects of capillary forces  and the 
microscopic origins of macroscopic cohesion $c$. Specifically we extend the results to different values of intergranular friction coefficient $\mu$, 
ranging from $0.05$ to $0.25$, and explore \emph{effective stress} ideas, which might provide microscopic predictions of the shear resistance and of macroscopic cohesion $c$ (if the Mohr-Coulomb criterion applies).
Note also that, thanks to the use of dimensionless variables, our results apply whatever the value of the liquid interfacial tension. 

In the following presentation of numerical results, data points are shown in figures with the same  symbol code,  consistently identifying
the five values of $\mu$ and the five values of liquid content $\Phi_L/\Phi_S$. This code is given in Table~\ref{tab:symbols}.
\begin{table*}[htb]
\centering
\caption{Choice of symbols used to identify intergranular friction coefficient $\mu$ and  liquid content $\Phi_L/\Phi_S$ in subsequent figures 
showing numerical simulation results. Colour codes for liquid contents are doubled by filling pattern.}
\label{tab:symbols}       
\begin{tabular}{|c||c|c|c|c|c|}
\hline
$\mu$ & $0.05$ & $0.09$ & $0.15$& $0.20$& $0.25$\\
\hline
Shape & squares & triangles & circles & stars & downward triangles\\
\hline
\hline
$\Phi_L/\Phi_S$ & $0.003$ & $0.007$ & $0.014$& $0.030$& $0.063$\\
\hline
Colour & red & blue & green & black & purple\\
Filling & filled & half (bottom) & hollow & crossed &half (top)\\
\noalign{\smallskip}\hline
\end{tabular}
\end{table*}
Thus numerical data pertaining \emph{e. g.}, to $\mu=0.09$ and $\Phi_L/\Phi_S=0.007$ are plotted as blue triangles with their bottom half filled.  
The choice of symbols in Fig.~\ref{fig:philphis}, \ref{fig:cexpnum} and \ref{fig:phiexpnum} for numerical data points abide by these codes. 
\subsection{The effective stress approach\label{sec:eff1}}
The effective stress approach to internal friction, as previously introduced in  papers dealing with wet grains~\cite{SCND09,SHNCD09,SJT16}, 
amounts to assuming that the friction law in quasistatic flow is the same as in the dry system subjected 
to the same \emph{effective stresses} $\tseff$, simply defined as $\tseff=\tscont$: the capillary forces 
acting on the grains are supposed to affect the contact network just like externally applied stresses. The Coulomb relation of the dry material thus applies to effective stress
components:
\be
\vert\sigma_{12}-\scap_{12}\vert= \mu^*_0 ( \sigma_{22}-\scap_{22}),
\label{eq:coulombeff}
\ee

\begin{figure}[htb!]
\centering
 \resizebox{0.92\columnwidth}{!}{\includegraphics{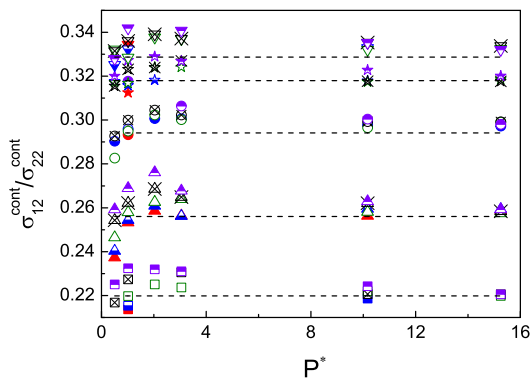}}
\caption{$\vert\scont_{12}\vert/\scont_{22}$ versus $P^*$ for all available values of  friction coefficient $\mu$ and meniscus volume $V_0$. 
Horizontal lines show $\mu$-dependent values of $\mu_0^*$. Symbols listed in Table~\ref{tab:symbols}.
\label{fig:coulombeff}}
\end{figure}

This prediction is directly tested on measuring contact stresses $\tscont = \tsigma-\tscap$ and plotting, in Fig.~\ref{fig:coulombeff},  ratios $\vert\scont_{12}\vert/\scont_{22}$ versus $P^*$ for
different intergranular friction coefficients $\mu$ and liquid contents $\Phi_L/\Phi_S$.  
The Coulomb relation for effective stresses, as written in  Eq.~\ref{eq:coulombeff}, involving the $\mu$-dependent internal friction coefficient $\mu_0^*$ of the 
dry material (as shown in Fig.~\ref{fig:mustarphimu}), provides a very good approximation of the material shear resistance in all studied cases. 
Larger discrepancies (between 5 and 10\%) tend to be observed for the smallest $P^*$ values, for small $\mu$ (as $\mu^*_0$ 
decreases and the relative importance of capillary effects increases), and for the largest meniscus volumes.

\begin{figure}[htb!]
\centering
(a)\resizebox{0.92\columnwidth}{!}{\includegraphics{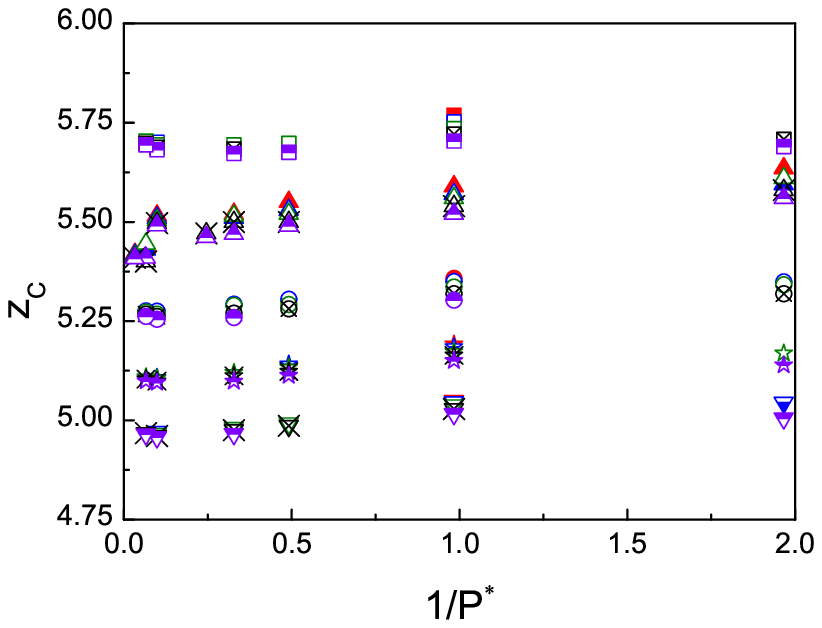}}\\
(b)\resizebox{0.92\columnwidth}{!}{\includegraphics{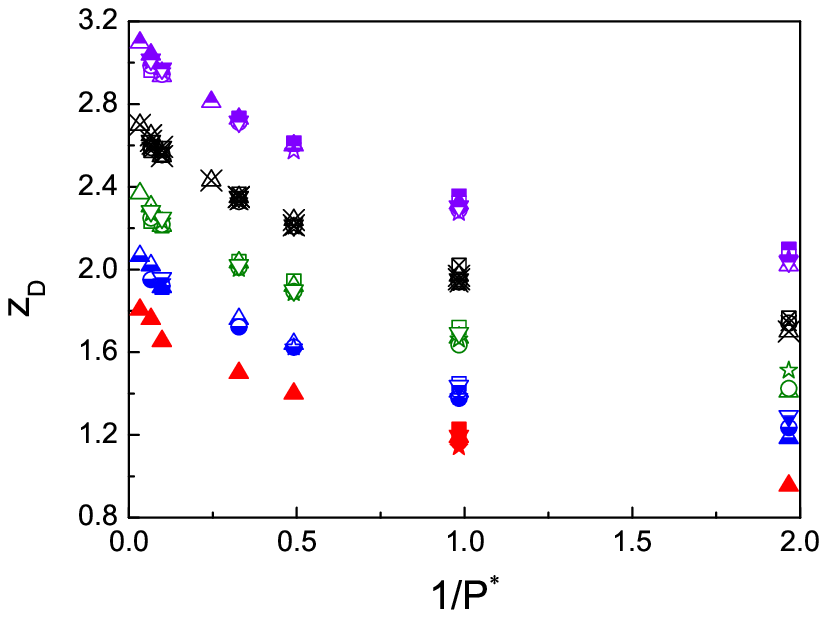}}\\
\caption{Variations with $P^*$ of (a) contact coordination number $\zc$ and (b) coordination number $\zd$ of 
distant interactions through liquid bridges, for different values of $\mu$ and $\Phi_L/\Phi_S$, encoded as in Table~\ref{tab:symbols}.
\label{fig:zzcip}}
\end{figure}

A possible clue to the remarkable success of the effective stress approach is that the changes in the contact network of the sheared material remain moderate between $P^*=0.5$ and the cohesionless limit of $P^*=\infty$. 
Thus, Fig.~\ref{fig:zzcip}a shows that contact coordination numbers hardly vary with $P^*$ and with liquid content, and are significantly influenced by friction coefficient $\mu$ (data points cluster by symbol shape, see Table~\ref{tab:symbols}).
This contrasts with the behavior of $\zd$ (Fig.~\ref{fig:zzcip}b), which varies between 1 and 3.2 for the investigated parameter range and is essentially determined by $\Phi_L/\Phi_S$ (encoded as colour and filling, see Table~\ref{tab:symbols}). $\zd$ also increases moderately with $P^*$.
\begin{figure}[htb!]
\centering
\resizebox{0.92\columnwidth}{!}{\includegraphics{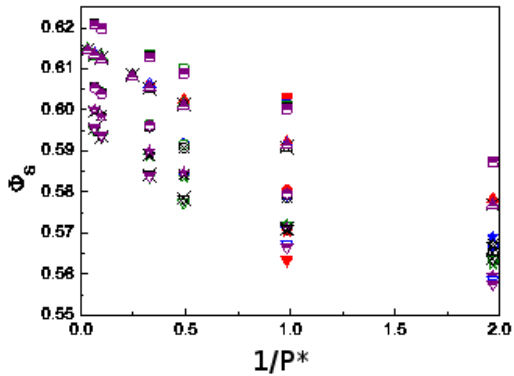}}
\caption{Variations of solid fraction $\Phi_S$  with $P^*$, for different values of $\mu$ and $\Phi_L/\Phi_S$ (symbols as in Table~\ref{tab:symbols}).
\label{fig:phisdiffmu}}
\end{figure}

As to the solid fraction, $\Phi_S$, Fig.~\ref{fig:phisdiffmu} shows its variation with $P^*$ in quasistatic shear flow for the different values of friction coefficient $\mu$, and liquid content $\Phi_L/\Phi_S$.
 A decreasing function of $\mu$, $\Phi_S$  varies moderately with reduced normal stress $P^*$, with a difference of 0.03 or 0.04 between the cohesionless limit of large $P^*$ and the lowest studied value $P^*=0.5$. 
It is very nearly independent of liquid content (as signaled by the clustering of points by symbol shape in Fig.~\ref{fig:phisdiffmu}).

To probe situations when the effective stress approach is more likely to fail by large amounts, 
one would need to study smaller $P^*$ values, for which  cohesion-dominated systems strongly depart from cohesionless ones  in their microstructure~\cite{KRC15,TKTPCR17}. 
As already mentioned, it then proves difficult to observe steady uniform shear flows, due to  strain localization in shear bands.
\subsection{Shear strength.\label{sec:predshear}}
The Coulomb condition for effective stresses, Eq.~\ref{eq:coulombeff}, is equivalent to 
this expression of apparent friction coefficient $\mu^* = \vert\sigma_{12}\vert/\sigma_{22}$:
\be
\mu^* = \mu^*_0  (1-\frac{\scap_{22}}{\sigma_{22}}) - \frac{\scap_{12}}{\sigma_{22}},
\label{eq:strength1}
\ee
which is satisfied in very good approximation. As a consequence of the sign of capillary forces, ratio $\scap_{22}/\sigma_{22} $ is negative and  is the cause of 
the considerable increase of shear resistance of the wet material compared to the dry one. The wet material is similar to the dry one, to which a larger normal
stress $\scont_{22}= \sigma_{22} - \scap_{22} >  \sigma_{22}$ is applied~\cite{KRC15}. 
On the other hand, as $\scap_{12}$ is positive while the total shear stress $\sigma_{12}$ is negative,
 the capillary force contributions to shear stress tends to decrease the material shear resistance.

We now discuss possible estimation schemes to predict the different capillary stress terms appearing in Eq.~\ref{eq:strength1}, so that the shear resistance of 
the wet material could be deduced from the internal friction coefficient $\mu^*_0$ as identified in the dry case, 
supplemented with the values of a few internal variables such as coordination numbers or fabric parameters.

Capillary stress component $\scap_{22}$, first, 
in view of the relatively small differences between normal stresses~\cite{KRC15}, is close to the capillary contribution to the average stress, $\cPcap$: we
explicitly checked that the relative difference never exceeds $3\%$ throughout our data set, remaining between 1 and 2\% in most cases.

Exploiting relation~\eqref{eq:PcapvsFN}, $\cPcap$ may be written as
\be
\cPcap = -\frac{Z^*\Phi_SF_0}{\pi a^2},
\label{eq:cest1}
\ee
$Z^*$ denoting some effective coordination number, larger than the  contact coordination number $\zc$, and smaller than the total coordination number $Z = \zc+\zd$,
such that the average capillary force for distant pairs,  $-F_d$, satisfies 
\be
\zd F_d + \zc F_0 = Z^*F_0.
\label{eq:defzstar}
\ee
Replacing 
$\scap_{22}$ by $\cPcap$ evaluated by \eqref{eq:cest1} in relation~\ref{eq:strength1}, one obtains:
\be
\mu^* = \mu^*_0  (1+\frac{\Phi_SZ^*}{\pi P^*}) - \frac{\scap_{12}}{\sigma_{22}}.
\label{eq:strength2}
\ee

Finally, a third level of approximation is obtained on discarding $\scap_{12}$ in \eqref{eq:strength2}:
\be
\mu^* = \mu^*_0  (1+\frac{\Phi_SZ^*}{\pi P^*}).
\label{eq:strength3}
\ee
We checked that relation~\eqref{eq:strength2}, involving the estimation of $\scap_{22}$ with approximate relation~\eqref{eq:cest1} still provides a very good prediction of stress ratio $\mu^*$. Neglecting the 
capillary force contribution to the shear stress as in~\eqref{eq:strength3}, as noted in~\cite{Chareyre},  leads to somewhat poorer predictions of the shear strength, 
 overestimated by 20--30\% in some cases. 
\begin{figure}[htb!]
\centering
\resizebox{0.92\columnwidth}{!}{\includegraphics{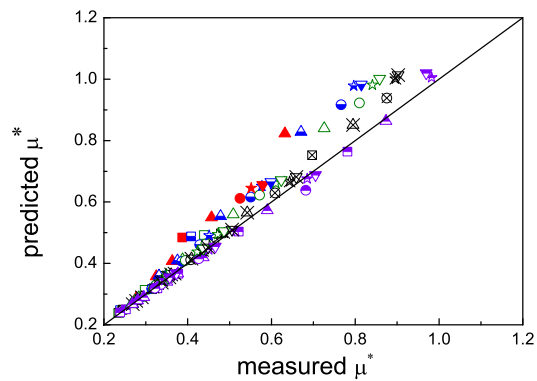}}
\caption{Predicted values of stress ratio $\mu^*$  using
 Eq.~\ref{eq:strength3}, versus measured ones, for all values of $\mu$ and $\Phi_L/\Phi_S$. Symbol codes as in Tab.~\ref{tab:symbols}. 
\label{fig:appdiff}}
\end{figure}
This is illustrated in Fig.~\ref{fig:appdiff}. It might be noted, too, that the error tends to grow as the liquid content decreases (from top-half filled, purple symbols, 
 to red, filled ones, see Table~\ref{tab:symbols}).
The importance and the origins of shear capillary stresses are further discussed in Sec.~\ref{sec:s12cap} below.  
We note, however, that expression~\ref{eq:strength3} might still provide an acceptable estimate of shear resistance in a smaller $P^*$ interval.

\begin{figure}[htb!]
\centering
\resizebox{0.92\columnwidth}{!}{\includegraphics{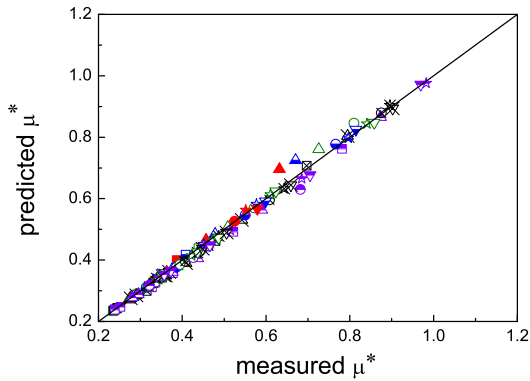}}
\caption{Predicted values of stress ratio $\mu^*$  using
 Eq.~\ref{eq:strength2}, versus measured ones, for all values of $\mu$ and $\Phi_L/\Phi_S$. Symbol codes as in Tab.~\ref{tab:symbols}.
\label{fig:s12pred}}
\end{figure}

Fig.~\ref{fig:s12pred}  shows  that a better prediction is  obtained in general on including the capillary shear stress, as in Eq.~\ref{eq:strength2}.
From the dry value $\mu^*_0\simeq 0.2$ obtained with $\mu=0.05$ and large $P^*$ up to the value of $\mu^*$, approaching one, corresponding to the system with  
$\mu = 0.25$, $\Phi_L/\Phi_S \simeq 0.063$ under $P^*=0.5$, all observed values of apparent friction coefficient are satisfactorily predicted from the
effective stress assumption.

In Sec.~\ref{sec:mccoh} below, we discuss predictions  of the Mohr-Coulomb cohesion, as defined and
measured in Sec.~\ref{sec:cexpnum} (see Fig.~\ref{fig:cphilphis}), corresponding to both estimates \eqref{eq:strength2} and \eqref{eq:strength3} of the shear 
resistance.
\subsection{Macroscopic cohesion\label{sec:mccoh}}
\begin{figure}[htb]
\centering
\resizebox{0.92\columnwidth}{!}{\includegraphics{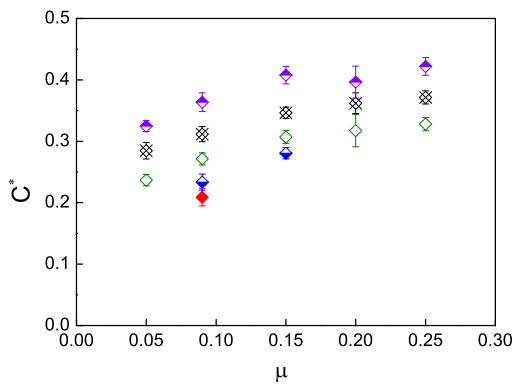}}
\caption{\label{fig:cstarmu} 
Reduced cohesion $c^*$ for different liquid contents $\Phi_L/\Phi_S$ (values encoded as in Table~\ref{tab:symbols}), 
versus intergranular friction coefficient $\mu$, as measured using  Eq.~\ref{eq:mohrcoulomb2}.
}
\end{figure}

On identifying the macroscopic cohesion by fitting 
a linear dependence to the variations of $\mu^* = \vert\sigma_{12}\vert/\sigma_{22}$ with $1/P^*$ 
(see Eq.~\ref{eq:mohrcoulomb2} and Fig.~\ref{fig:cexpnum}), 
the resulting values, for different intergranular friction coefficients, are shown in Fig.~\ref{fig:cstarmu}.
Reduced cohesion $c^*$  increases with $\mu$ and with the liquid content, typically by 30 to 50\% over the explored interval  of each variable. 
A similar moderate increase of macroscopic cohesion with liquid content or saturation through the interval corresponding 
to the pendular regime is reported in a number of 
DEM studies and experimental measurements on wet bead assemblies~\cite{SECC06,RYR06,SCND09}.

Relation~\ref{eq:strength3} is equivalent  to the Mohr-Coulomb criterion~\eqref{eq:mohrcoulomb2} and thus provides an expression of $c^*$:
\be
c^* = \frac{\Phi_S Z^*\mu^*_0}{\pi}.
\label{eq:cohesion3}
\ee
One recognizes  of course the classical Rumpf formula, as written in \eqref{eq:Rumpf} 
(with a slightly modified definition of the appropriate coordination number, see Eq.~\ref{eq:defzstar}).
A constant, $P^*$-independent value of $c^*$ might be obtained, as a good approximation in view of the 
limited variations of coordination numbers and solid fraction (see Figs.~\ref{fig:zzcip} and \ref{fig:phisdiffmu}), on averaging expression \ref{eq:cohesion3} 
over some range of $P^*$. Fig.~\ref{fig:cphilphis} compares  this ``Rumpf formula" estimate of the cohesion, averaged for  $P^*\ge 2.3$, to its value 
obtained by fitting the Mohr-Coulomb relation~\eqref{eq:mohrcoulomb2} to the data  in the same range.

\begin{figure}[htb]
\centering
\resizebox{0.92\columnwidth}{!}{\includegraphics{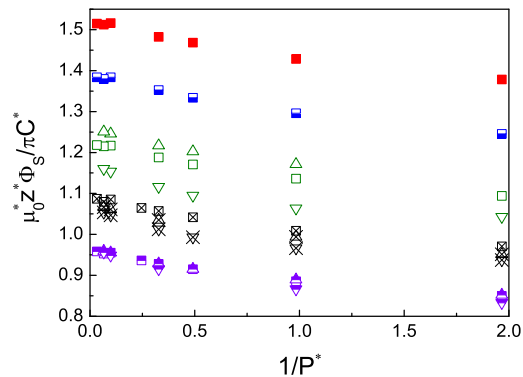}}
\caption{\label{fig:cohcomp} 
Ratio of reduced cohesion $c^*$  as predicted by  Eq.~\ref{eq:cohesion3} to its directly identified value, through a fit of data to Eq.~\ref{eq:mohrcoulomb2}, versus $1/P^*$, 
for different  friction coefficients and liquid contents. Symbol codes as in Table~\ref{tab:symbols}.}
\end{figure}

We find that $F_d$, as defined in~\eqref{eq:deffd},
depends on liquid content, more than on $\mu$. It hardly varies with $P^*$, 
and remains  between $0.48\times F_0$ and $0.6\times F_0$ for $P^*\ge 2$, 
with  maxima near $0.7\times F_0$ for the lowest $P^*$ values.
The approximation $Z^*=\zc+0.5\times \zd$ [see Eq.~\ref{eq:defzstar}] may be used in Eq.~\ref{eq:cohesion3}. 
Fig.~\ref{fig:cohcomp} compares the resulted estimated $c^*$ values (equal to the ``Rumpf formula'' data of Fig.~\ref{fig:cphilphis})
 to the measured ones, shown in Fig.~\ref{fig:cstarmu}.
(As a check on an estimate of macroscopic cohesion, the values shown in Fig.~\ref{fig:cohcomp} should only matter for $1/P^* < 0.5$, for which the Mohr-Coulomb criterion applies. 
For smaller $P^*$, the data points on the figure merely show the variations of capillary normal stress $\scap_{22}$). 

Despite the satisfactory agreement already recorded in Fig.~\ref{fig:cphilphis} for $\mu=0.09$ and $\Phi_L/\Phi_S \ge 0.03$, 
Fig.~\ref{fig:cohcomp} shows that the accuracy of the prediction
of Eq.~\ref{eq:mohrcoulomb2} may significantly deteriorate for lower liquid contents, especially for small friction coefficients (i.e., from Table~\ref{tab:symbols}, for data points shown as squares and triangles). 

In Ref.~\cite{KRC15}, it was pointed out that estimate~\eqref{eq:cohesion3} applied satisfactorily 
to the experimental results of Pierrat \emph{et al.}~\cite{PAC98} and Richefeu \emph{et al.}~\cite{RYR06}\footnote{In Ref.~\cite{RYR06}  a theoretical estimate of cohesion $c$ is used, differing from Eq.~\ref{eq:cohesion3} by a factor of 3/2, but it may be argued that 
the uncertainty in measured values of $c$ in~\cite{RYR06} is such that our prediction is acceptable as well. 
The estimate of~\cite{RYR06}  is reproduced in~\cite{SCND09,AFP13}.}. 
One advantage of prediction~\eqref{eq:cohesion3} is its simplicity~\cite{Reply}. 
In practice coordination numbers could be obtained from accurate X-ray tomography observations~\cite{KOH04}, or perhaps inferred from elastic moduli~\cite{iviso3}.

Beyond the simple estimation of \eqref{eq:cohesion3}, 
the success of the effective stress approach justifies the definition of a cohesion in a generalized sense, as the contribution of
capillary forces to shear resistance, in addition to the effects of the dry material internal friction coefficient $\mu^*_0$, which still relates elastic-frictional components of
the contact stress tensor, $\tscont$. This leads to a possibly $P^*$-dependent reduced macroscopic cohesion $c^*$ 
given by:
\be
c^*_g(P^*) =  \frac{\mu^*_0\Phi_S Z^*}{\pi} - \frac{a^2\scap_{12}}{F_0},
\label{eq:cohgen}
\ee
in which the Rumpf estimate~\eqref{eq:cohesion3} gets corrected by the second term.
\begin{figure}[htb]
\centering
\resizebox{0.92\columnwidth}{!}{\includegraphics{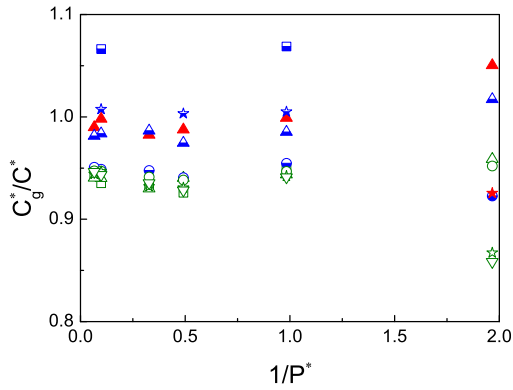}}
\caption{\label{fig:cohgen} 
Ratio of  generalized cohesion $c^*_g$  as predicted by  Eq.~\ref{eq:cohgen} to its  value identified by a Mohr-Coulomb fit of the data,  
versus $1/P^*$, for different  friction coefficients and small liquid contents. Symbol codes as in Table~\ref{tab:symbols}.}
\end{figure}
Fig.~\ref{fig:cohgen}, analagously to Fig.~\ref{fig:cohcomp}, compares this generalized cohesion to the value obtained by a Mohr-Coulomb fit to the data.
Prediction~\eqref{eq:cohgen} achieves satisfactory accuracy in the range of parameters for which \eqref{eq:cohesion3} fails. For larger liquid contents, as
the simpler prediction of \eqref{eq:cohesion3} is fairly accurate, we observe no improvement on using Eq.~\eqref{eq:cohgen} instead. 
Expression~\eqref{eq:cohgen} is compared to experimental data at low liquid content (and to numerical ones with the 
experimentally relevant value $\mu=0.09$ of the intergranular friction coefficient) in Fig.~\ref{fig:cphilphis}.

Interestingly, Shen \emph{et al.}, in a DEM study of triaxial compression of wet bead assemblies~\cite{SJT16}, also 
showed that the deviator part of the capillary stress tensor contributes 
negatively to the macroscopic cohesion.


In order to  relate $c^*_g$, as  defined by Eq.~\ref{eq:cohgen}, or the shear strength, as 
written in Eq.~\ref{eq:strength1}, to characteristic state variables, 
Sec.~\ref{sec:s12cap} below investigates the micromechanical origins of capillary shear stress $\scap_{12}$. 
 
\section{Capillary shear stresses.\label{sec:s12cap}}
As the direct contribution of capillary forces to shear stress,  $\scap_{12}$, should not be neglected in general,
we now investigate its dependence on control parameters (Sec.~\ref{sec:s12capdep}), 
and its microscopic origins (Sec.~\ref{sec:s12capor}), highlighting the different roles of
capillary forces in contacting and distant pairs. The specific contribution of the latter, which, to a large extent, explains the dependence of shear strength and cohesion on liquid content $\Phi_L/\Phi_S$, 
is related in Sec.~\ref{sec:s12capdis} to geometric and mechanical anisotropy parameters of the liquid bridge network. 
Finally, the complete relation of $\scap_{12}$  to internal variables such as coordination numbers and anisotropy parameters is briefly discussed in Sec.~\ref{sec:s12capconc}. 
\subsection{Dependence on $P^*$, $\mu$ and $\Phi_L/\Phi_S$ \label{sec:s12capdep}}
\begin{figure}[htb]
\centering
\resizebox{0.92\columnwidth}{!}{\includegraphics{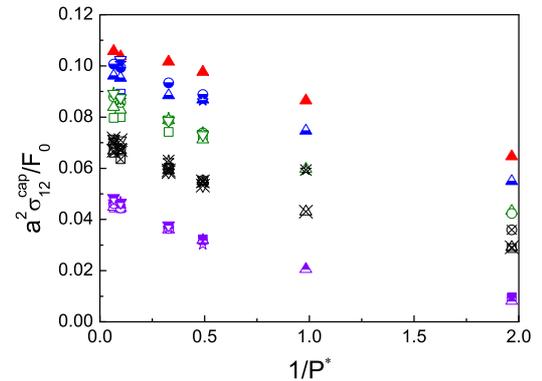}}
\caption{\label{fig:scaps} 
$\scap_{12}$, in units of $F_0/a^2$, versus $1/P^*$ for different liquid contents and 
intergranular friction coefficients (symbols as in Table~\ref{tab:symbols}). 
}
\end{figure}

As previously noted, due to the attractive nature of capillary forces, the sign of $\scap_{12}$ is  opposite to that of $\scont_{12}$, 
or of $\sigma_{12}$: one has $\scap_{12}>0$.
Fig.~\ref{fig:scaps} shows the variations of 
$\scap_{12}$ through the explored parameter range.
$a^2\scap_{12}/F_0$  -- the correction, according to Eq.~\ref{eq:cohgen}, to the reduced cohesion $c^*$ estimated with Eq.~\ref{eq:cohesion3} -- 
depends on $\Phi_L/\Phi_S$ and on $P^*$, and hardly varies with $\mu$ (data points, with the code of Table~\ref{tab:symbols}, 
order by colour and filling, but not by symbol shape).
It strongly decreases for growing $\Phi_L$ and very notably increases with $P^*$.  

$a^2\scap_{12}/F_0$ values  range from nearly 0 (for large $\Phi_L/\Phi_S$  and small $P^*$) to about $0.1$ (for small  $\Phi_L/\Phi_S$  and  large $P^*$), 
while reduced cohesion $c^*$, as shown in Figs.~\ref{fig:cphilphis} and \ref{fig:cstarmu}, grows with $\Phi_L/\Phi_S$ and with $\mu$, varying in range $0.2\le c^*\le 0.4$. 
The correction introduced by the capillary shear stress in the evaluation of the shear strength, or on estimating a macroscopic cohesion, is thus particularly important for small
liquid content and for small intergranular friction coefficient, as could be observed in Figs.~\ref{fig:appdiff} and \ref{fig:cohcomp}.
\subsection{Role of contacts and distant interactions \label{sec:s12capor}}
Capillary stresses can be split into the contribution of liquid bridges joining  grains in contact, and another one
due to bridges joining noncontacting pairs. We respectively denote  those terms with superscripts ``cap,c'', and ``cap,d'':
\be
\scap_{12} = \scapc_{12}  +\scapd_{12}.
\label{eq:s12capcd}
\ee
Those two terms, in the sum of Eq.~\ref{eq:s12capcd}, are of opposite signs, with $ \scapc_{12}>0$ and  $\scapd_{12}<0$.  
Thus, on writing $\sigma_{12} = \scont_{12} + \scapc_{12}  +\scapd_{12}$, the three terms of the right-hand side alternate in sign. 

\begin{figure}[htb]
\centering
\resizebox{0.92\columnwidth}{!}{\includegraphics{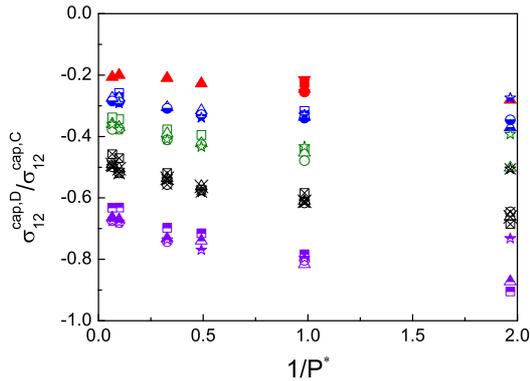}}
\caption{\label{fig:scapcd} 
Ratio of contributions of distant pairs and of contacting ones to capillary shear stress, $\scapd_{12}/\scapc_{12}$, 
versus $1/P^*$ for different liquid contents and intergranular friction coefficients. Symbol codes as in Table~\ref{tab:symbols}.
}
\end{figure}

Fig.~\ref{fig:scapcd} plots (negative) ratio $\scapd_{12}/\scapc_{12}$ versus $P^*$ throughout the explored parameter range. 
 The contact contribution, $ \scapc_{12}$, is larger in magnitude and imposes its sign to the sum. It dominates for small $\Phi_L/\Phi_S$, 
 but the influence of the contribution  of  distant interactions grows with saturation, and  both 
contributions more nearly compensate as the upper limit of the pendular regime is approached.

\begin{figure}[htb]
\centering
(a)\resizebox{0.92\columnwidth}{!}{\includegraphics{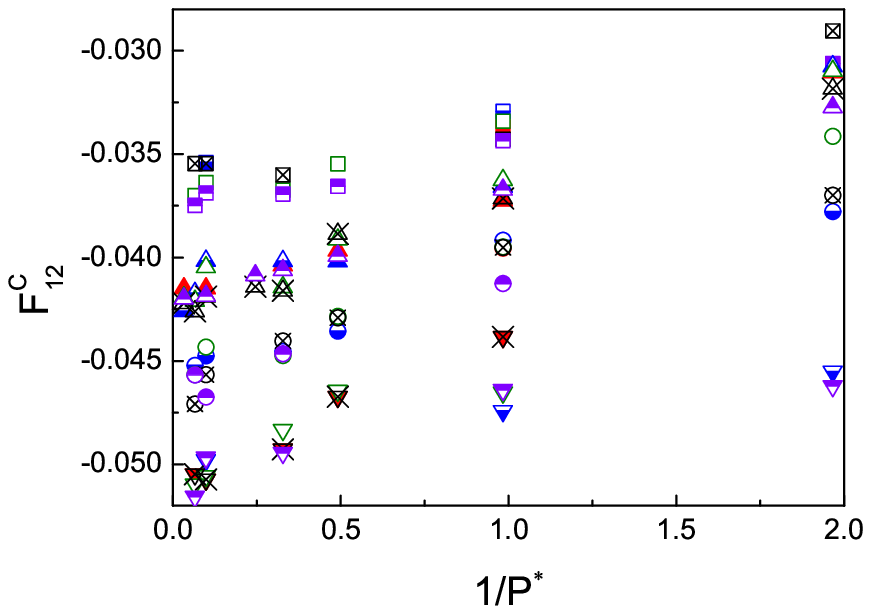}}\\
(b)\resizebox{0.92\columnwidth}{!}{\includegraphics{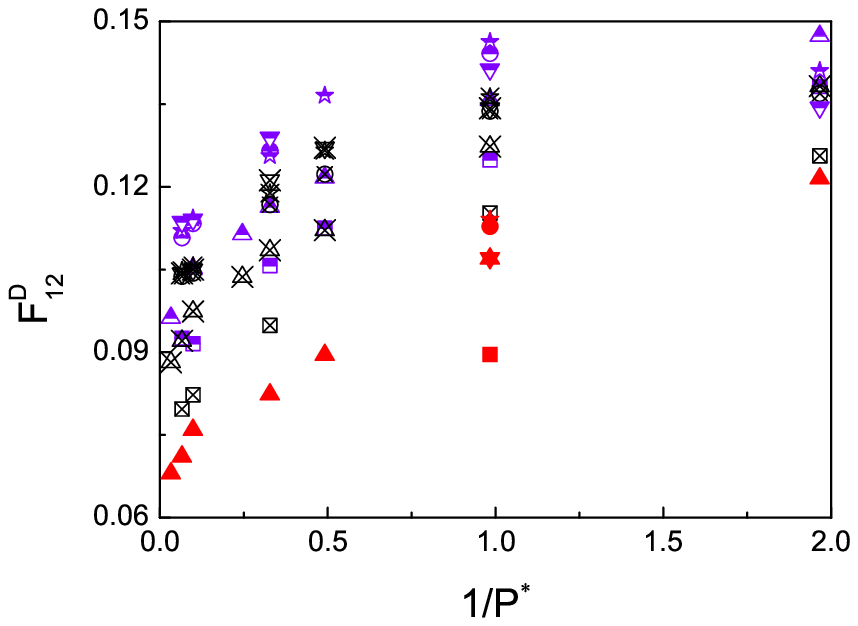}}
\caption{\label{fig:fabric} 
Fabric parameters  $\fccis$ for contacts (a),  and $\fdcis$ for distant interactions (b) versus $1/P^*$. Symbol codes as in Table~\ref{tab:symbols}.
}
\end{figure}

From the second moments of the distribution of normal unit vectors ${\bf n}$ (the direction of the capillary force, carried by the
line of centres), we define the second-order fabric tensors, $\ww{F}^c$ and $\ww{F}^d$,  respectively associated  with contacting pairs and with distant pairs joined by a meniscus.
Fabric anisotropy is well known to be related to  stress anisotropy~\cite{RB89,PR08b,SHNCD09,AR12,RDAR12}.
The components of $\ww{F}^c$ and $\ww{F}^d$ relevant to the evaluation of shear stresses $\sigma_{12}$ are 
(subscripts $c$ or $d$ respectively indicating averages over contacts and over distant interacting pairs)
\be
\ba
\fccis&=\ave{n_1 n_2}_c &\ \ \mbox{for contacts}\\
\fdcis&=\ave{n_1 n_2}_d&\ \ \mbox{for distant interactions.}
\ea
\label{eq:defF12}
\ee
Both fabric parameters are plotted in Fig.~\ref{fig:fabric}, versus $1/P^*$, for the different values of $\mu$ and liquid content $\Phi_L/\Phi_S$. 
$\fccis$  is primarily influenced by friction coefficient $\mu$, and is hardly affected by liquid content. 
It remains roughly constant (between -0.036 and -0.047, depending on $\mu$) through the investigated interval of $P^*$. 
The variations of $\fdcis$ with $P^*$ exhibit a slight, but notable decrease as $P^*\ge 1$.  $\fdcis$ grows with liquid content and (to a lesser extent) with $\mu$. 
$\fdcis$ is larger than $\vert\fccis\vert$, reaching values in the range 0.07--0.14. 

Fabric parameter $\fdcis$, as noted in Ref.~\cite{KRC15}, is positive, while $\fccis$ is negative -- in agreement with the signs of $\scapc_{12}$  and $\scapd_{12}$. 
This may be understood on noting that, in the macroscopic shear flow (Fig.~\ref{fig:sheartest}), contacts tend to be more numerous for orientations ${\bf n}$  in the compression quadrants (for which $n_1n_2<0$) than in the extension quadrants (for which $n_1n_2>0$). 
Conversely, as direction  ${\bf n}$ 
rotates within the shear flow and enters the extension quadrants, gaps tend to open between contacting pairs of grains, which still interact for distances below $D_0$  and contribute positively to $\fdcis$. Liquid bridges between non-contacting grains were observed in \cite{KRC15} 
to survive over strain intervals of order 1. 
It may be noted that 
Ref.~\cite{SHNCD09}, which contains a DEM study of triaxial compression on wet beads, with a similarly hysteretic model for meniscus formation,  reports that
the fabric anisotropy of the liquid bridge network is considerably smaller than the fabric anisotropy of the contact  network, implying that the liquid bridges joining non-contacting grains contribute with opposite sign. 
Capillary forces in contacting pairs are all equal to $-F_0$ in our model, and their centers, neglecting elastic deflections, are 
separated by distance $a$. Consequently, from expression \eqref{eq:stress} of shear stress, one has:
\be
\frac{a^2\scapc_{12}}{F_0} = -\frac{3\zc \Phi_S}{\pi} \fccis,
\label{eq:s12capcdecomp}
\ee
\subsection{Shear stress due to distant interactions\label{sec:s12capdis}}
For distant interactions, one may also define the ${\bf n}$  dependent average force  (the average capillary force between distant pairs sharing the same intercentre line orientation ${\bf n}$), 
as $\fa({\bf n}) $, 
as well as the ${\bf n}$-dependent average intercentre distance $\ld({\bf n})$. 
The average intercentre distance for non-contacting bead pairs interacting  through a liquid bridge is denoted as $l_0$ (with $a\le l_0\le a+D_0$). 
$l_0$ varies very little with $P^*$ and $\mu$, it is essentially dependent on liquid content (or $V_0$), and grows from $1.025$ to nearly $1.06$ over the explored range of $\Phi_L$. 

\begin{figure}[htb]
\centering
\resizebox{0.92\columnwidth}{!}{\includegraphics{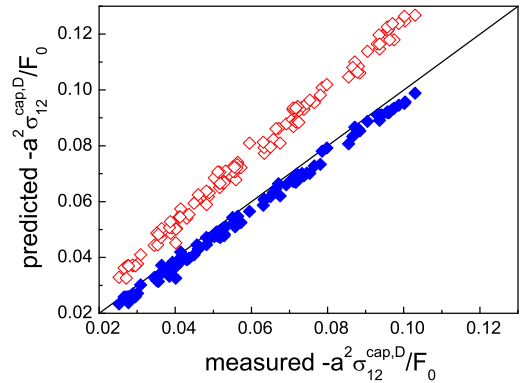}}
\caption{\label{fig:scapdapp} 
Predicted values of $-a^2\scapd_{12}/F_0$, using ~\eqref{eq:s12capddecomp},  versus measured ones, for the complete numerical data set. 
Open red diamonds: prediction with the first term of \eqref{eq:s12capddecomp} (fabric anisotropy). Full blue ones: prediction with the first two terms of \eqref{eq:s12capddecomp} (fabric and force anisotropies).}
\end{figure}

Standard relations~\cite{RB89,PR08b,AzRa2014}, based on the first non-trivial term in the expansion of even anisotropic functions of ${\bf n}$  in spherical harmonics,
directly relate shear stress contributions to fabric, force and distance anisotropy parameters. 
Defining, through integrations on the unit sphere $S$ with the differential solid angle $d\Omega$, force and distance anisotropy parameters as
\be
\ba
\ffdcis &= \frac{-1}{4\pi F_d}\int_ S \fa({\bf n}) n_1n_2 d\Omega \\
\lldcis &=  \frac{1}{4\pi l_0}\int_ S \ld({\bf n}) n_1n_2 d\Omega,
\ea
\label{eq:deffl12}
\ee
one may write
\be
\frac{a^2\scapd_{12}}{F_0} \simeq -\frac{3\Phi_S \zd }{\pi}\frac{F_d}{F_0}\frac{l_0}{a} \left[\fdcis + \ffdcis + \lldcis\right],
\label{eq:s12capddecomp}
\ee
as an approximation. 
As shown in Fig.~\ref{fig:scapdapp}  relation \eqref{eq:s12capddecomp} is quite accurate. 
The term involving $ \lldcis$ proves negligible (of relative order $10^{-2}$).

Comparing relations \ref{eq:s12capcdecomp}  and  \ref{eq:s12capddecomp}, the reason for which $\vert \scapd_{12} \vert <  \scapc_{12}$ (see Fig.~\ref{fig:scapcd}), despite the opposite inequality 
for fabric parameters ($ \vert \fccis\vert>\fdcis$, see Fig.~\ref{fig:fabric}) is that product $l_0 \zd F_d$ is significantly smaller than $a\zc F_0$. Furthermore, the force anisotropy parameter $\ffdcis$ is negative in  \eqref{eq:s12capddecomp}, 
reducing the magnitude of $\scapd_{12}$ (one observes $\ffdcis <0$ and $\vert\ffdcis\vert <\fdcis$). 

To explain the  sign of $\ffdcis$, it should be noted that non-contacting grains in interaction tend to be closer to one another, whence a larger attractive force, for force orientations ${\bf n}$ 
in the compression quadrants ($n_1n_2<0$). One thus have $\lldcis>0$ and  $\ffdcis<0$ [it should be recalled that $\ffdcis$, as defined in~\eqref{eq:deffl12},  
is the moment of ${\bf n}$-dependent average force distribution $\fa({\bf n})$ \emph{normalized by the global average} $-F_d$]. 
$\ffdcis$ does not vary much with $P^*$, $\mu$ and liquid content, with typical values between -0.02 and -0.04.
\subsection{Estimating capillary shear stress \label{sec:s12capconc}}
We thus obtain a very good estimate of $\scap_{12}$ on summing Eqs.~\ref{eq:s12capcdecomp} and \ref{eq:s12capddecomp} (neglecting the last term):
\be
\frac{a^2\scap_{12}}{F_0} \simeq -\frac{3\Phi_S}{\pi}\left[ \zc\fccis +\frac{l_0}{a} \zd \frac{F_d}{F_0} \left(\fdcis + \ffdcis  \right)\right].
\label{eq:estfinals12c}
\ee
\begin{figure}[htb]
\centering
\resizebox{0.92\columnwidth}{!}{\includegraphics{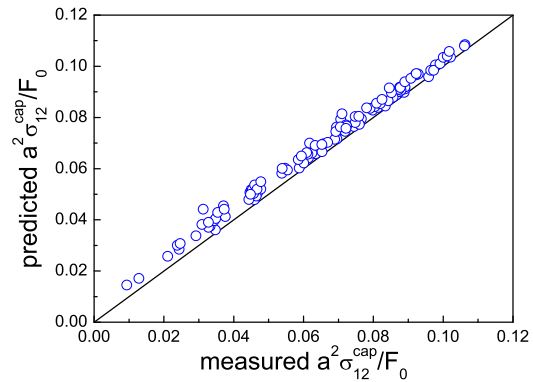}}
\caption{\label{fig:scapapp} 
$\scap_{12}$, in non-dimensional form, as predicted with Eq.~\ref{eq:estfinals12c},  versus its measured value, for the complete numerical data set.  
}
\end{figure}
This is checked in Fig.~\ref{fig:scapapp}.
The capillary force contribution to the shear stress, $\scap_{12}$, is thus related to a small number of internal variables characterizing the networks formed 
by the intergranular contacts and network and the additional liquid bonds connecting non-contacting grains.

Another relation of  macroscopic cohesion to fabric and force anisotropy parameters was proposed by  Radjai and Richefeu~\cite{RR09}, based on a somewhat different approach: these authors did not exploit the effective stress idea, but related the macroscopic friction coefficients of both the wet material and the dry one to their structural and force anisotropy parameters, such that the cohesion is expressed in terms of the differences of those variables induced by the capillary forces. 

In \eqref{eq:estfinals12c}, the three terms of the right-hand-side are of alternating signs and decreasing magnitudes. 
In general, none of them is negligible: distant interactions, by reducing the contribution of capillary forces in contacts, notably influence the capillary shear stress; 
and the effect of the rather large fabric anisotropy onto $\scapd_{12}$ gets reduced by the one of  force anisotropy. 
 The dependence of the macroscopic cohesion and shear strength on the liquid content are largely due to the effect of $\scapd_{12}$, 
 the distant interaction contribution to  $\scap_{12}$.  Cohesion $c$ increases with $\Phi_L$ because $\vert\scapd_{12}\vert$ increases, thereby decreasing 
$\scap_{12}$, which is  the term reducing the macroscopic cohesion (Eq.~\ref{eq:cohgen}).

 \section{Summary and discussion\label{sec:conc}}
We now  recall the salient results of the paper. 

First,  experimental measurements for shear strength and solid fraction in  steady uniform quasistatic shear flow of a model material, made of polystyrene beads wet by silicone oil, are reported. 
A Mohr-Coulomb criterion applies well for  values of reduced normal stress $P^*$ larger than a few units, but tends to overestimate the shear strength for lower  $P^*$. It involves the static macroscopic friction coefficient $\mu^*_0$ of the dry material, and a macroscopic cohesion $c$,
which grows with liquid content $\Phi_L/\Phi_S$ in the pendular
regime. In non-dimensional form, $c^*=a^2c/F_0$  varies between $0.2$ and $0.4$. 

Then, results of DEM simulations are shown to agree quantitatively with experimental ones, provided the appropriate value $\mu\simeq 0.09$ is given to the intergranular friction coefficient, as identified from the 
macroscopic properties (internal friction coefficient $\mu^*_0$ and solid fraction $\Phi_S$) of the dry material.

Further use of numerical simulations,  for different values of $\mu$, enabled investigations of the microscopic origins of macroscopic shear resistance and cohesion, and assessments of the performance of  existing approaches and prediction schemes. 
The effective stress approach, assuming that the capillary part of the stress tensor acts
onto the contact network like an externally applied stress, as expressed by relation~\eqref{eq:coulombeff} and checked in Fig.~\ref{fig:coulombeff}, 
proves remarkably efficient (although not exact). In general,  the contribution of capillary forces to shear stress should not be ignored, as it significantly improves the prediction of the shear resistance (as apparent on comparing Figs.~\ref{fig:appdiff} and \ref{fig:s12pred}). Capillary stress components $\scap_{22}$ (see Eq.~\ref{eq:cest1}) and $\scap_{12}$ (see Eq.~\ref{eq:estfinals12c})  are both related to coordination numbers and to characteristics of the force network (involving anisotropy parameters for shear stress), so that it is possible to express shear strength and cohesion with simple
predictive formula involving the macroscopic friction coefficient of the dry material. The ``Rumpf formula", Eq.~\ref{eq:cohesion3}, 
for macroscopic cohesion $c$, is based on the effective stress approach in which $\scap_{12}$ is neglected. 
It is approximately correct for large enough liquid content within the pendular range ($\Phi_L/\Phi_S\ge 0.03$), but despite the 
dependence of the wet coordination number on $\Phi_L$, proves unable (as shown in Figs.~\ref{fig:cphilphis} and \ref{fig:cohcomp})  
to capture the liquid content dependence of the cohesion for smaller saturations. 
The contribution of $\scap_{12}$ to shear strength is the largest, in relative terms, for small liquid contents, 
especially for small values of friction coefficient $\mu$, as the Rumpf term is proportional to macroscopic friction coefficient $\mu^*_0$ 
(a growing function of $\mu$, see Fig.~\ref{fig:mustarphimu}). 
The cohesion  increase with $\Phi_L$ originates in the decrease of  capillary stress component $\scap_{12}$, which contributes negatively to cohesion. 
This effect is due to the opposite fabric orientations between contacts and liquid bridges joining noncontacting particles.   

The success of the effective stress approach is likely related to the relative insensitivity of the contact network structure on capillary effects: solid fraction $\Phi_S$ (Fig.~\ref{fig:phisdiffmu}), contact coordination number $\zc$ (Fig.~\ref{fig:zzcip}a), and
contact fabric parameter $\fccis$ (Fig.~\ref{fig:fabric}) are essentially determined by $\mu$, hardly depend on $\Phi_L/\Phi_S$, and exhibit little (or moderate) variations with $P^*$. On the other hand, $\zd$, the coordination number
of distant interactions, and $\scapd_{12}$, their contribution to shear stress (which  explains the dependence of $\scap_{12}$ on liquid content), 
are nearly independent of $\mu$, and vary considerably with $\Phi_L/\Phi_S$ and with $P^*$. 

The investigations of the influence of the liquid content are one original aspect of the paper. 
Although the results shown in Fig.~\ref{fig:cphilphis} are quite encouraging, one may wish to explore to what extent  they are sensitive to the spatial distribution of the liquid phase, which is admittedly rather crudely modeled in the present numerical study. 
An obvious, but technically challenging, extension of the present work would be to investigate the
capillary effects at saturations beyond the pendular regime. This requires some continuum  mechanics modeling of the liquid phase configuration, 
which requires technically challenging numerical models well beyond the reach of simple DEM. Some interesting attempts involve a  Lattice-Boltzmann treatment of the interstitial fluid phases~\cite{DeRiRa15,RiRaDe16}. 
On the experimental side, rheological measurements could be usefully supplemented by investigations of liquid morphologies
through microtomography~\cite{KOH04,HER05,MKOH12,BPVLDB13b}. Although accurate determinations of contacts in a grain pack are very difficult (see~\cite{ASSS04,ASS05} for measurements and~\cite{iviso1} for a discussion with comparisons to simulations), liquid bridges, due to the larger scales
involved, are easier to observe, and, knowing the distance-dependent capillary force, some of the characteristic state variables of the capillary force network (coordination and fabric parameters as involved, e.g., in Eq.~\ref{eq:estfinals12c}) used in Sec.~\ref{sec:s12cap} might be experimentally accessible.

The properties of the loose  structures stabilized by cohesion observed under small $P^*$~\cite{KBBW03,GiRoCa08,TKTPCR17}  are also worth studying, as regards their gradual deformation and collapse under varying loads. Such aspects (as does, already, the small $P^*$ dependence of density and contact network
observed in this paper) escape the effective stress approach, and their study 
 would enable a more global assessment of its range of applicability. Such 
 an exploration  could not be pursued here in the range $P^*\sim 0.1$, as localization phenomena preclude observations of homogeneous critical states in steady quasistatic shear flow. The
conditions of occurrence of such strain localization (which may entail fluid distribution inhomogeneities~\cite{MKOH12}), is
an issue of great practical consequences,  and would also deserve  systematic investigations. 

\subsection*{Acknowledgements}
We are grateful to David Hautemayou and C\'edric M\'ezi\`ere for technical helps with the measurements. This research was partially funded by 
Agence Nationale de la Recherche (Grant No. ANR-16-CE08-0005-01).
\bibliographystyle{unsrt}
\bibliography{refs}
\end{document}